\documentclass[a4paper]{jpconf}  
\usepackage[dvips]{graphicx}
\usepackage{dcolumn}   
\usepackage{bm}        
\usepackage{amssymb}   
\usepackage[utf8]{inputenc}
\usepackage{mathrsfs}
\usepackage{amsmath}
\usepackage{hyperref}
\usepackage{textcomp}
\usepackage{subfigure}
\usepackage{color}

\begin{document}

\title{Resonance phenomena in the $\varphi^8$ kinks scattering}

\author{E Belendryasova$^{1}$ and V A Gani$^{1,2}$ }

\address{$^1$National Research Nuclear University MEPhI (Moscow Engineering Physics Institute), 115409 Moscow, Russia}

\address{$^2$National Research Center Kurchatov Institute, Institute for Theoretical and Experimental Physics, 117218 Moscow, Russia}

\ead{vagani@mephi.ru}

\begin{abstract}
We study the scattering of the $\varphi^8$ kinks with power-law asymptotics. We found two critical values of the initial velocity, $v_\mathrm{cr}^{(1)}$ and $v_\mathrm{cr}^{(2)}$, which separate different regimes of the kink-antikink collision. At the initial velocities $v_\mathrm{in}< v_\mathrm{cr}^{(1)}$ kinks can not collide due to repulsive force between them. At $v_\mathrm{in}>v_\mathrm{cr}^{(2)}$ the kinks escape to spatial infinities after one collision. In the range $v_\mathrm{cr}^{(1)}\le v_\mathrm{in}\le v_\mathrm{cr}^{(2)}$ we observed kinks capture and formation of their bound state. Besides that, at these initial velocities we found resonance phenomena --- escape windows.
\end{abstract}

\section{Introduction}

Scattering of one-dimensional topological defects (kinks) is very interesting and fast developing area. Many field theoretical models in (1+1)-dimensional space-time have kink-like solutions. The most famous are models with polynomial potentials, such as $\varphi^4$, $\varphi^6$ \cite{GaKuLi,MGSDJ}, $\varphi^8$ \cite{GaLeLi}--\cite{Belendryasova.arXiv.2017}, and ones with non-polynomial potentials --- modified sine-Gordon, double sine-Gordon \cite{GaKuPRE,Gani.arXiv.2017.dsg}, sinh-deformed $\varphi^4$ \cite{Bazeia.arXiv.2017.sinh,Bazeia.arXiv.2017.sinh.conf}, etc., \cite{Bazeia.PRD.2002,Bazeia.PRD.2006}. Note that models with two or more fields, which have topologically non-trivial solutions, are also actively studied recently \cite{GaKsKu01}--\cite{GaLiRaconf}.

The (1+1)-dimensional field-theoretical models are very important for numerous applications. For example, models with polynomial potentials are often used in the description of domain walls, phase transitions in condensed matter and in the early Universe, etc. Models with non-polynomial potentials, such as sine-Gordon, modified sine-Gordon, and double sine-Gordon, are used to describe disoriented chiral condensate \cite{Gani.YaF.1999,Gani.YaF.2001}, some processes in ferro-magnets, Josephson contacts, in some nematic liquid crystals, and so on. One of the classical examples is that the dynamics of planar domain walls can be reduced to the kink-(anti)kink interaction \cite{GaKu.SuSy.2001}. Topologically non-trivial field configurations can be responsible for formation of the primordial black holes in the early Universe \cite{GaKiRu,GaKiRu.conf}.

In this short letter we present our first results of numerical simulation of the kink-antikink collisions withing the $\varphi^8$ model. Kinks of this model have one power-law and one exponential asymptotics. In our numerical experiments the kink and antikink are turned to each other by power-law tails. This leads to long-range interaction between the kinks, which substantially affect their scattering. Here we address only phenomenology of the kink-antikink scattering. A detailed study of mechanisms, which lead to these phenomena, will be a subject of a separate publication.

\section{The $\varphi^8$ model}

We consider a field-theoretical model with a real scalar field $\varphi(x,t)$ in two-dimensional space-time with the Lagrangian
\begin{equation}
\label{eq:lagrangian}
\mathcal{L} = \frac{1}{2}\left(\frac{\partial \varphi}{\partial t}\right)^2-\frac{1}{2}\left(\frac{\partial \varphi}{\partial x}\right)^2-V(\varphi),
\end{equation}
where potential has the form
\begin{equation}\label{eq:potential}
V(\varphi)=\varphi^4(1-\varphi^2)^2.
\end{equation}
The equation of motion for the field $\varphi$ is partial differential equation of second order:
\begin{equation}
\label{eq:eqmo}
\frac{\partial^2\varphi}{\partial t^2} - \frac{\partial^2\varphi}{\partial x^2} + \frac{dV}{d\varphi} = 0.
\end{equation}
The static kinks can be obtained from the first order ordinary differential equation
\begin{equation}\label{eq:red_stat_eqmo}
\frac{d\varphi}{dx}=\pm\sqrt{2V(\varphi)}
\end{equation}
The solution with $\varphi(-\infty)<\varphi(+\infty)$ is called 'kink', while 'antikink' stands for the solution with $\varphi(-\infty)>\varphi(+\infty)$. Nevertheless, in many cases below we call both of them simply 'kinks'.

\begin{figure}[h!]
\centering\includegraphics[width=0.5\linewidth]{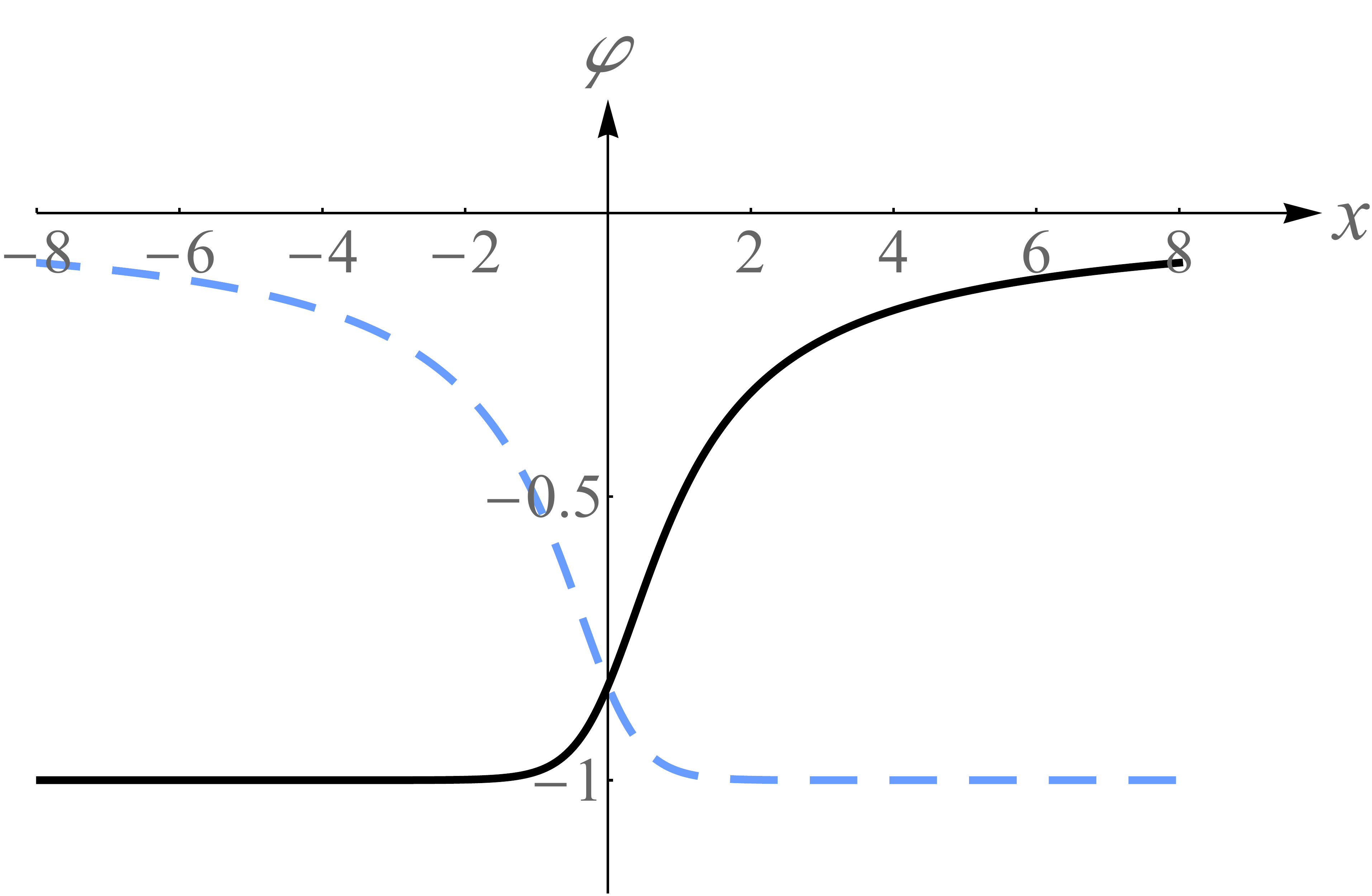}
\caption{Kink $(-1,0)$ and antikink $(0,-1)$ of the $\varphi^8$ model.}
\label{fig:kinks}
\end{figure}

In a model of $\varphi^8$ the integration of equation \eqref{eq:red_stat_eqmo} leads to implicitly defined kinks:
\begin{equation}
\label{eq:kinks}
\pm 2\sqrt{2}\: x = -\frac{2}{\varphi}+\ln\frac{1+\varphi}{1-\varphi},
\end{equation}
see figure \ref{fig:kinks}
The kinks \eqref{eq:kinks} belong to the topological sectors $(-1,0)$ and $(0,1)$, while antikinks --- to the sectors $(1,0)$ and $(0,-1)$. Asymptotics of all kinks are the following:
\begin{equation}\label{eq:kink1_asymp_minus}
\varphi_{(-1,0)}(x)\approx-1+\frac{2}{e^2}\: e^{2\sqrt{2}\: x},\quad \mbox{at} \ x\to -\infty,
\end{equation}
\begin{equation}\label{eq:kink1_asymp_plus}
\varphi_{(-1,0)}(x)\approx-\frac{1}{\sqrt{2}\: x},\quad \mbox{at} \ x\to +\infty,
\end{equation}
\begin{equation}\label{eq:kink2_asymp_plus}
\varphi_{(0,1)}(x)\approx-\frac{1}{\sqrt{2}\: x}, \quad \mbox{at} \ x\to -\infty,
\end{equation}
\begin{equation}\label{eq:kink2_asymp_minus}
\varphi_{(0,1)}(x)\approx 1-\frac{2}{e^2}\: e^{-2\sqrt{2}\: x}, \quad \mbox{at} \ x\to +\infty.
\end{equation}
As one can see from \eqref{eq:kink1_asymp_minus}--\eqref{eq:kink2_asymp_minus}, each kink has one exponential and one power-law asymptotics.

\section{Resonance phenomena in the kinks scattering in the sector $(-1,0,-1)$}

\begin{figure}[h!]
\centering\includegraphics[width=0.5\linewidth]{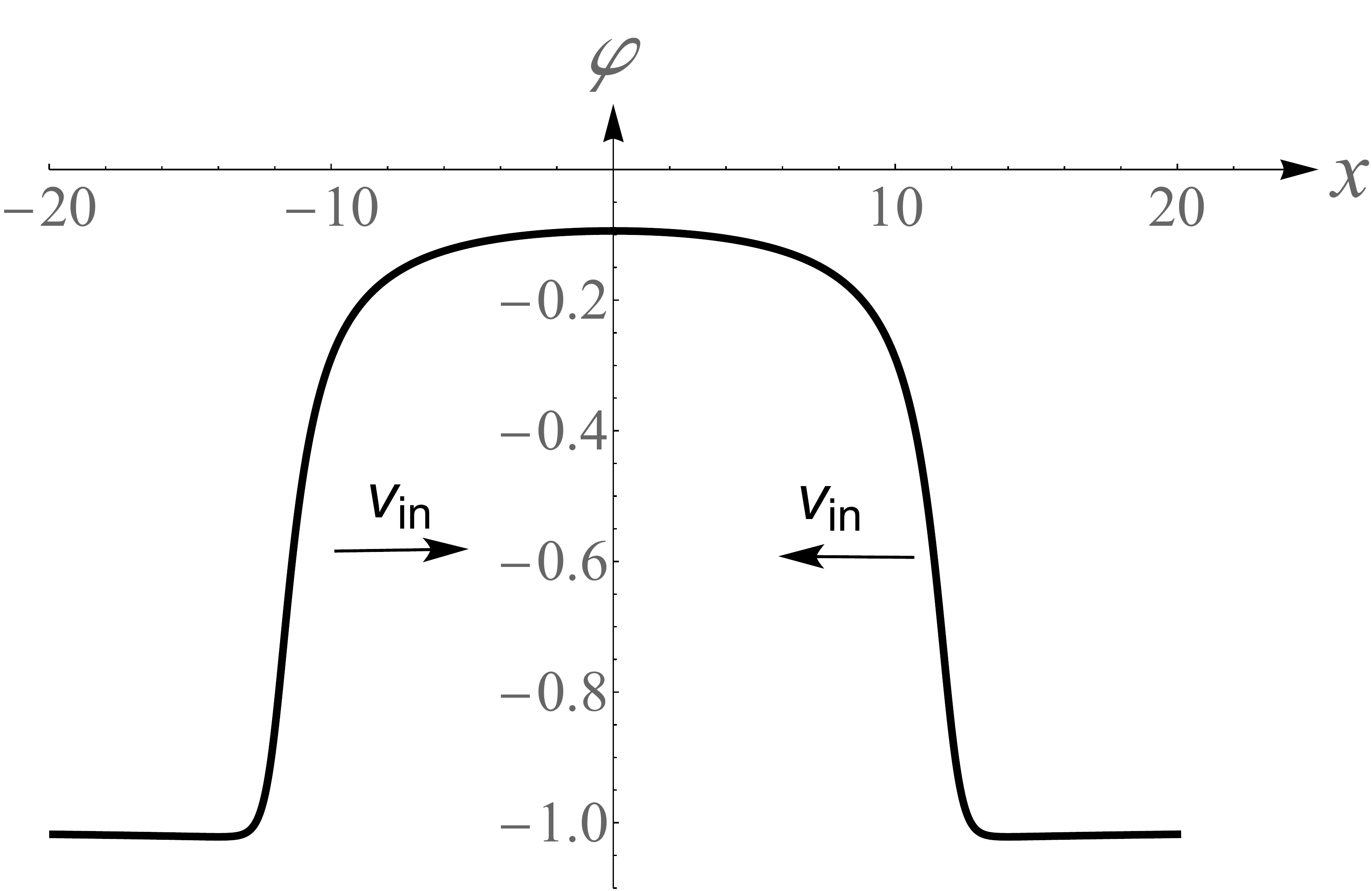}
\caption{The initial condition of the type of \eqref{eq:kink2_asymp_minus}.}
\label{fig:init_cond}
\end{figure}

We study collisions of kinks with power-law tails, i.e.\ in our numerical simulation the two colliding kinks are faced to each other by algebraic tails. Let us take the initial condition in the form of the kink $\varphi_{(-1,0)}$ and the antikink $\varphi_{(0,-1)}$, which are initially separated by distance $2\xi$ and are moving towards each other:
\begin{equation}\label{eq:incond}
\varphi(x)=\varphi_{(-1,0)}\left(\frac{x+\xi}{\sqrt{1-v_\mathrm{in}^2}}\right)+\varphi_{(0,-1)}\left(\frac{x-\xi}{\sqrt{1-v_\mathrm{in}^2}}\right),
\end{equation}
see figure \ref{fig:init_cond}. The initial velocity of each kink is $v_\mathrm{in}$. In our numerical simulation we used the initial kink-antikink distance $2\xi=30$.

We solved the second order partial differential equation \eqref{eq:eqmo} using the explicit finite-difference scheme,
\begin{equation}
\frac{\partial^2\varphi}{\partial t^2} = \frac{\varphi_j^{k+1}-2\varphi_j^k+\varphi_j^{k-1}}{\delta t^2}, \quad \frac{\partial^2\varphi}{\partial x^2} = \frac{\varphi_{j+1}^k-2\varphi_j^k+\varphi_{j-1}^k}{\delta x^2},
\end{equation}
on a grid with steps $\delta t=0.008$, $\delta x=0.01$. 

\begin{figure}[h!]
\subfigure[\:\ Scattering at velocity $v_\mathrm{in}=0.0806<v_\mathrm{cr}^{(1)}$]{
\begin{minipage}{0.5\linewidth}
\centering\includegraphics[width=0.88889\linewidth]{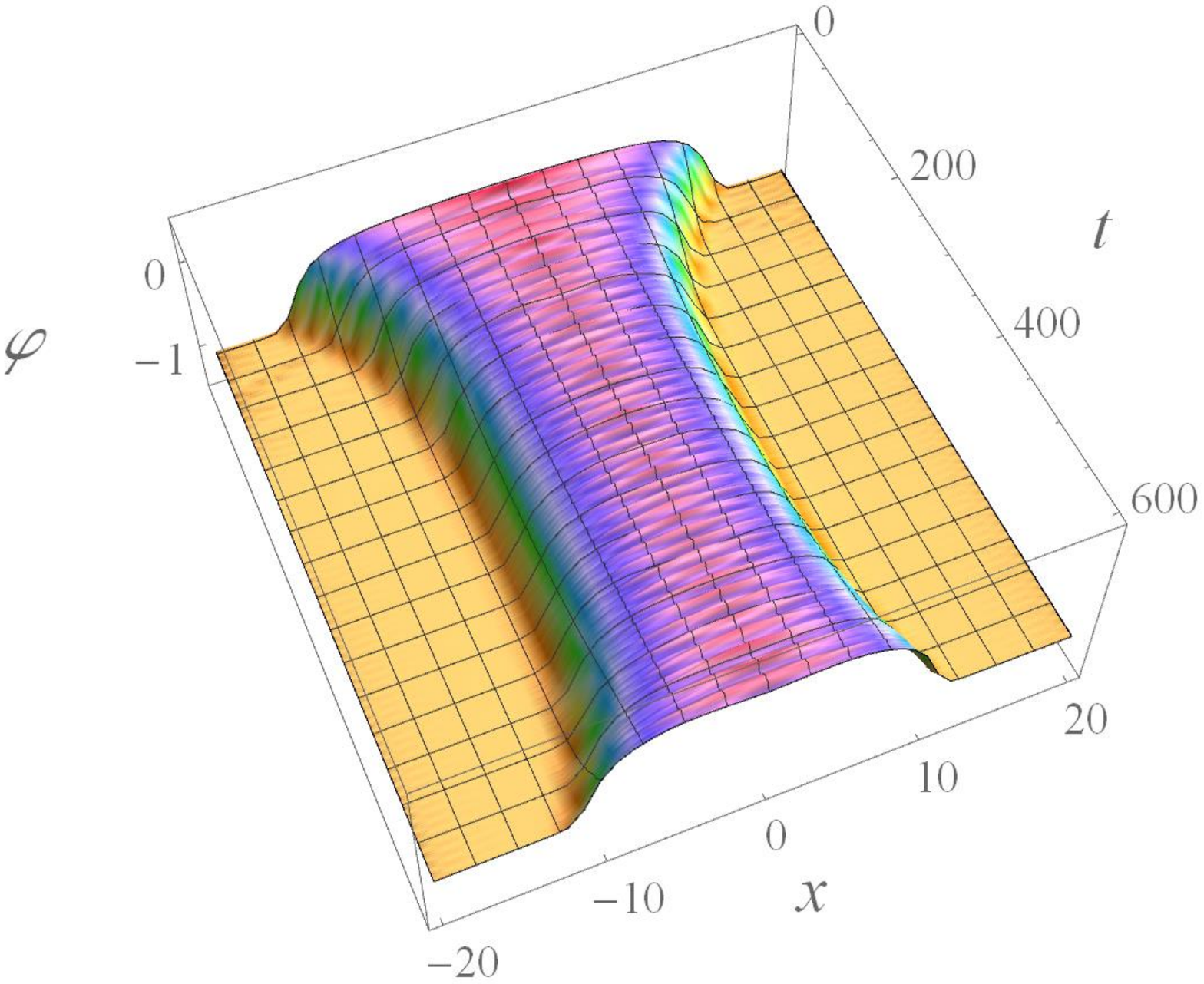}
\end{minipage}\label{fig:without_col}
}
\subfigure[\:\ Scattering at velocity $v_\mathrm{in}=0.2000>v_\mathrm{cr}^{(2)}$]{
\begin{minipage}{0.5\linewidth}
\centering\includegraphics[width=0.88889\linewidth]{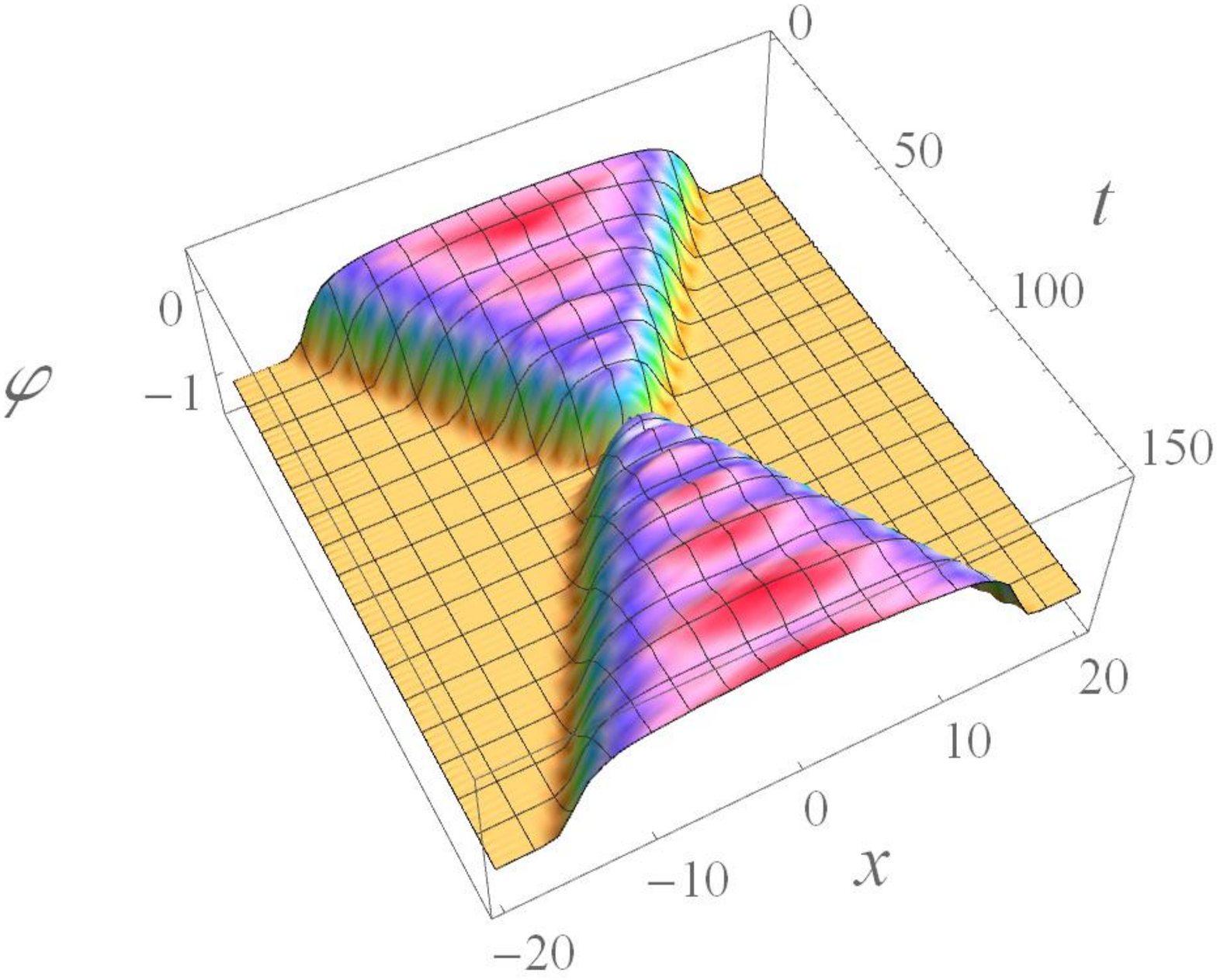}
\end{minipage}\label{fig:one_col}
}
\caption{Space-time picture of the kinks scattering at $v_\mathrm{in}<v_\mathrm{cr}^{(1)}$ and $v_\mathrm{in}>v_\mathrm{cr}^{(2)}$.}
\end{figure}

\begin{figure}[h!]
\subfigure[\:\ Bion formation, $v_\mathrm{in}=0.14510$]{
\begin{minipage}{0.45\linewidth}
\centering\includegraphics[width=\linewidth]{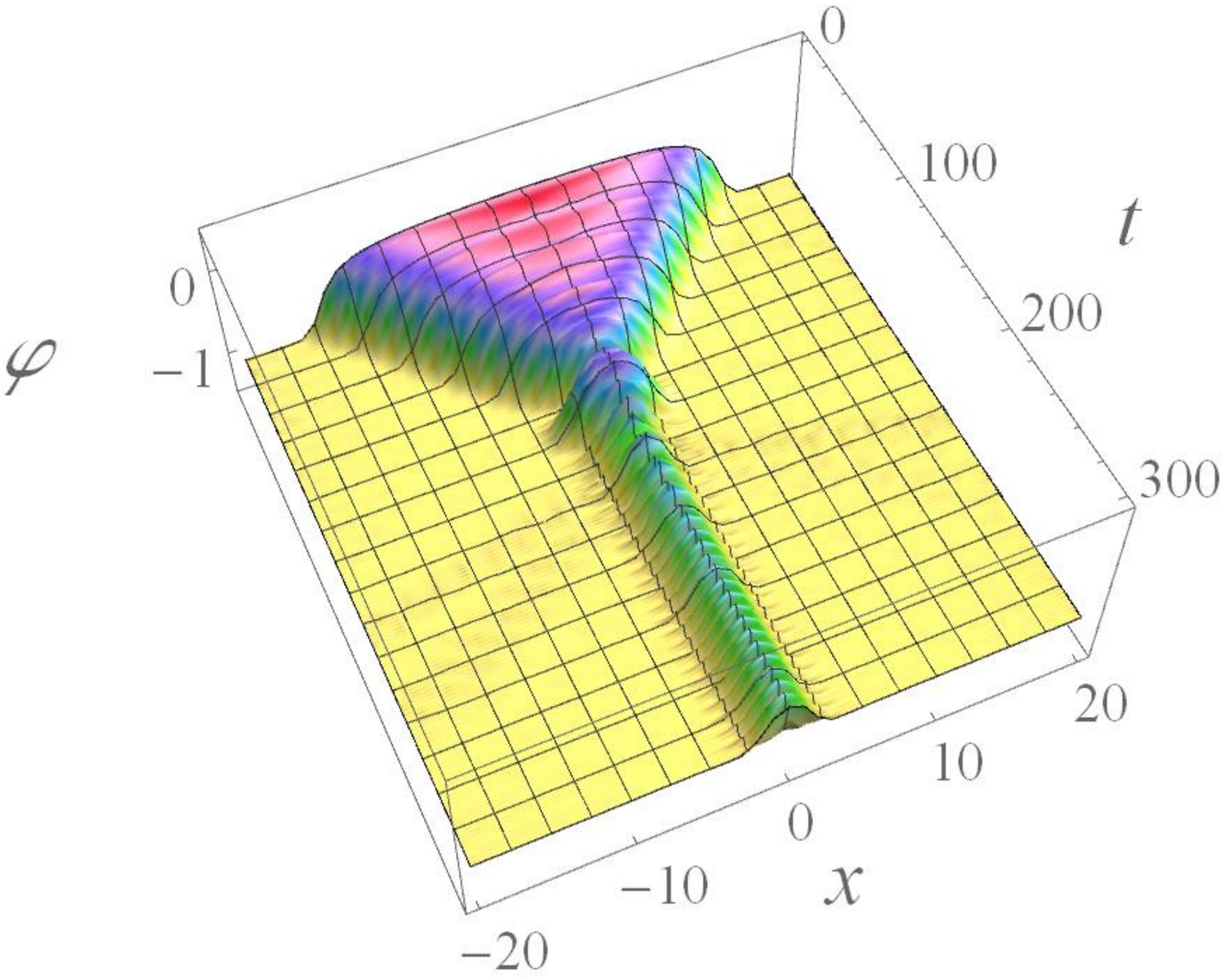}
\end{minipage}
\hspace{10mm}
\begin{minipage}{0.45\linewidth}
\centering\includegraphics[width=\linewidth]{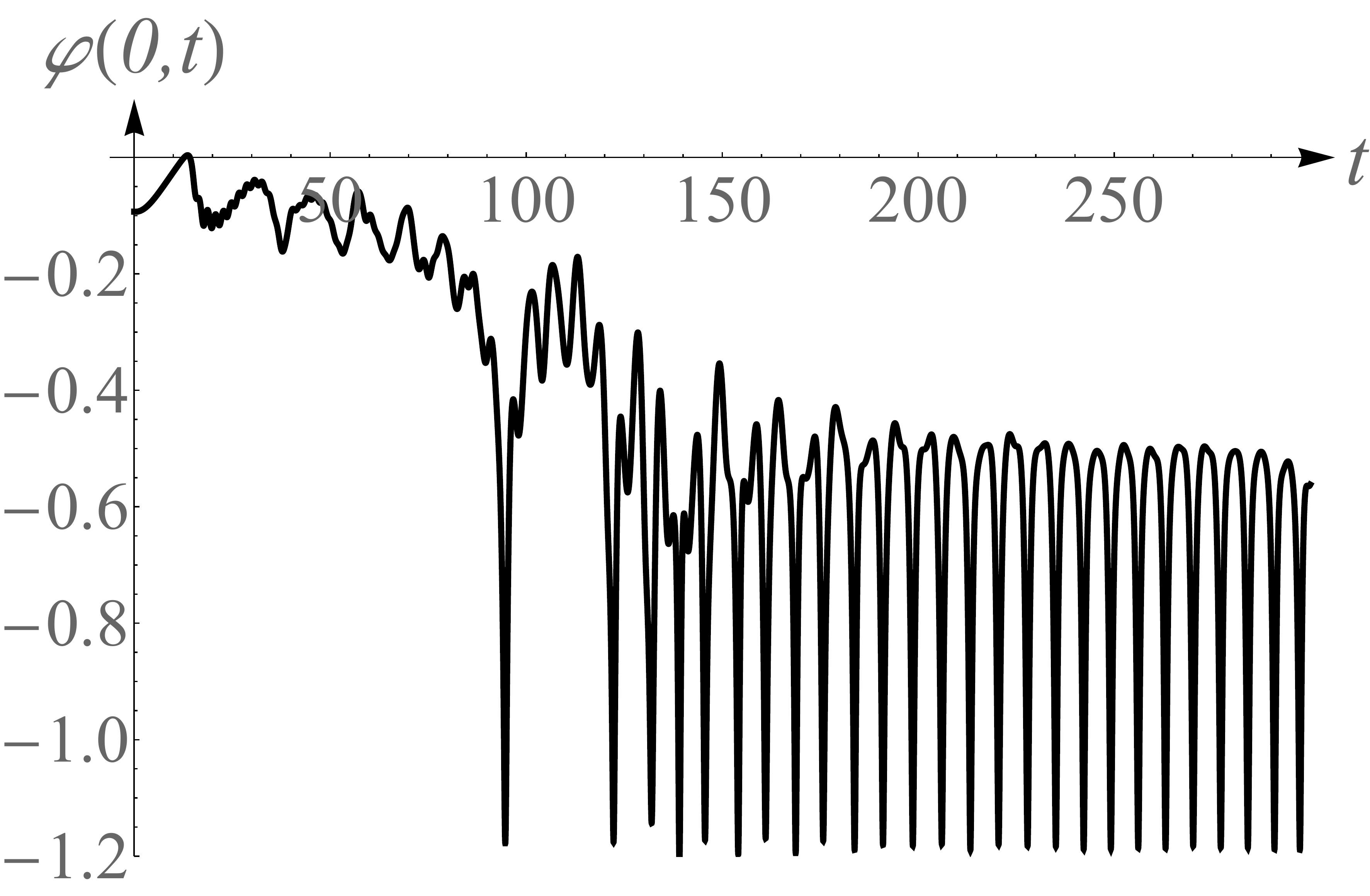}
\end{minipage}\label{fig:bion}}
\subfigure[\:\ Two-bounce window, $v_\mathrm{in}=0.14724$]{
\begin{minipage}{0.45\linewidth}
\centering\includegraphics[width=\linewidth]{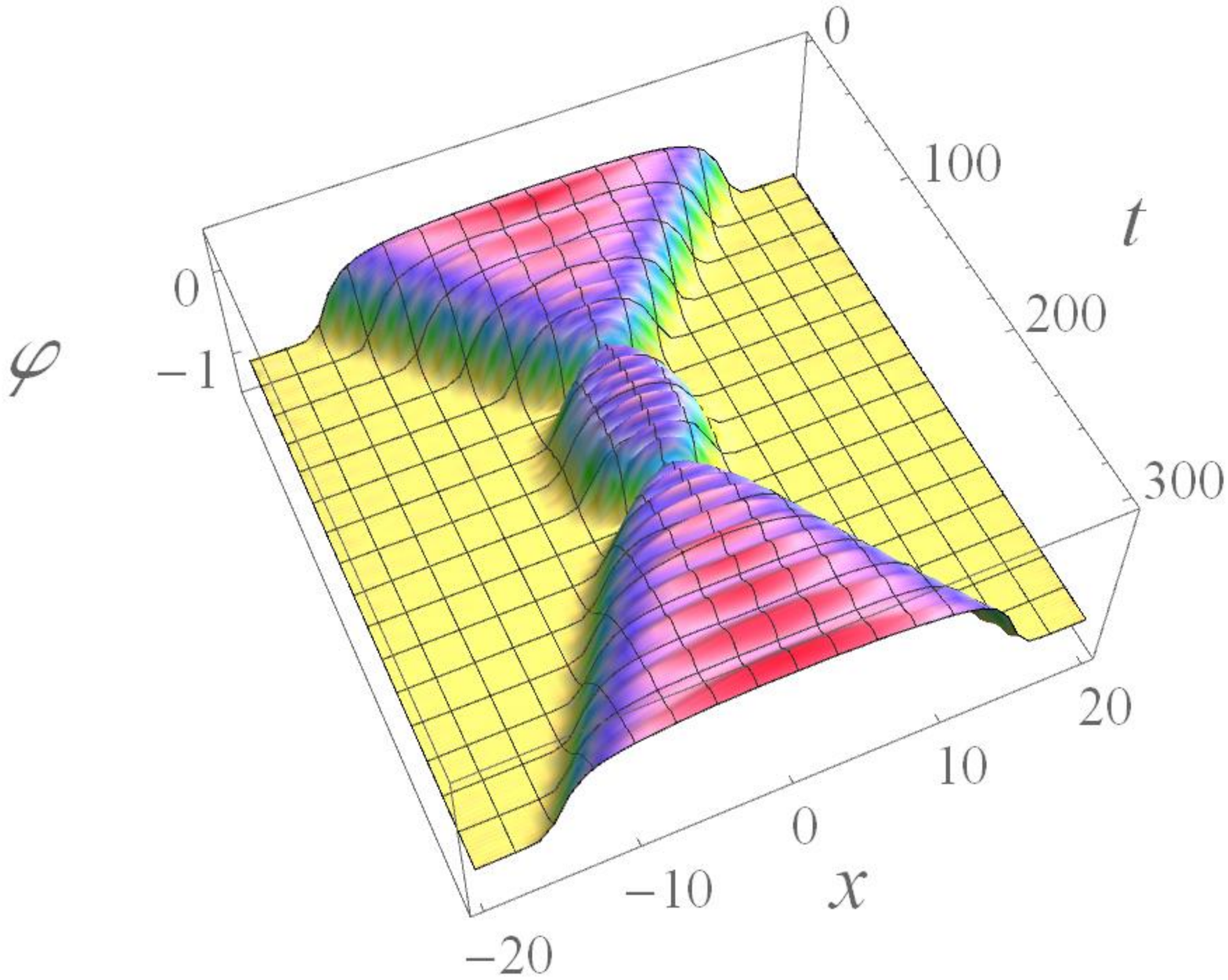}
\end{minipage}
\hspace{10mm}
\begin{minipage}{0.45\linewidth}
\centering\includegraphics[width=\linewidth]{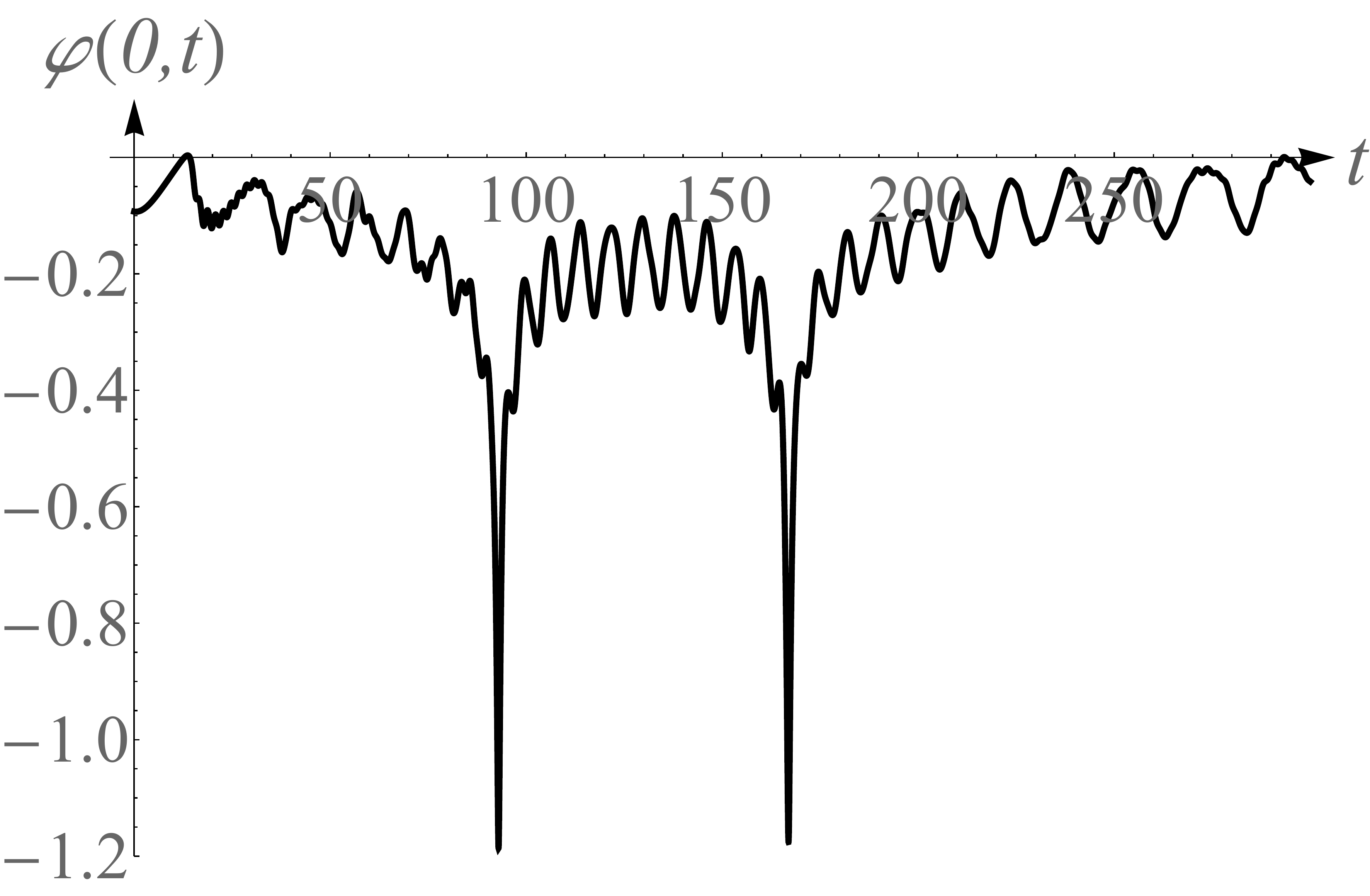}
\end{minipage}\label{fig:2win}
}
\caption{Space-time picture of the kinks scattering at $v_\mathrm{cr}^{(1)}\le v_\mathrm{in}\le v_\mathrm{cr}^{(2)}$.}
\end{figure}

Depending on the initial velocity, the kink-antikink collision looks very different. We have found two critical values of the initial velocity, $v_\mathrm{cr}^{(1)}\approx 0.08067$ and $v_\mathrm{cr}^{(2)}\approx 0.14791$, which separate different regimes of the kinks collisions.
\begin{enumerate}
\item At $v_\mathrm{in}<v_\mathrm{cr}^{(1)}$ the kink and the antikink do not collide because they can not overcome mutual repulsion. The kinks approach, stop, and then move away from each other, see figure \ref{fig:without_col}.
\item At $v_\mathrm{in}>v_\mathrm{cr}^{(2)}$ the kinks collide and escape to infinities, this scenario is shown in figure \ref{fig:one_col}.
\item At $v_\mathrm{cr}^{(1)}\le v_\mathrm{in}\le v_\mathrm{cr}^{(2)}$ we observed
\begin{enumerate}
\item kinks capture and formation of their bound state (a bion), see figure \ref{fig:bion};
\item two-bounce escape windows, see figure \ref{fig:2win};
\item multi-bounce escape windows --- three-and four-bounce escape windows.
\end{enumerate}
\end{enumerate}

The escape windows appear due to resonant energy exchange between translational mode (kinetic energy) and vibrational mode(s) of the colliding kinks. For more detailed discussion of this phenomenon see, e.g., \cite{GaKuLi,GaLeLi}.

An attentive reader can ask, how can it be that kinks repel, and at the same time they can form a bound state? At the moment we do not have definite answer. This situation is a challenge for us, and we will study the forces between the $\varphi^8$ kinks with power-law tails in the near future.

\section{Conclusion}

We have investigated numerically the scattering of the $\varphi^8$ kinks with power-law asymptotics. Such kinks are highly interactive in the sense that the force between them falls off algebraic with distance \cite{Radomskiy.JPCS.2017}. This, in turn, leads to long-range interaction between kinks --- they 'feel' each other at much bigger distances than kinks with exponential tails.

The results of our numerical simulation are very interesting. On the one hand, we observed repulsive forces between kinks. On the other hand, we have found resonant structures --- escape windows, which are typical to models with attractive kink-antikink interaction. Hence, this our study opens wide prospects for future work.
\begin{itemize}
\item First, it is interesting to investigate the force between kink and antikink as a function of kink-antikink distance. This force seems to be repulsive at large distances, and it could be attractive at small distances.
\item Second, kinks of the $\varphi^8$ model with the potential \eqref{eq:potential} have no vibrational mode, which could accumulate energy and could be responsible for the appearance of the escape windows. If so, where is the kinetic energy stored between the first and the second kinks collisions within the two-bounce escape window?
\end{itemize}
These and other issues will be the subject of a future study.

\section{Acknowledgements}

This work was performed using resources of the NRNU MEPhI high-performance computing center. The research was supported by the MEPhI Academic Excellence Project under contract \textnumero~02.a03.21.0005, 27.08.2013.


\section*{References}
\begin{thebibliography}{99}

\bibitem{GaKuLi} V.~A.~Gani, A.~E.~Kudryavtsev, and M.~A.~Lizunova, \href{https://doi.org/10.1103/PhysRevD.89.125009}{{Phys.~Rev.~D} {\bf 89}, 125009 (2014)} [\href{https://arxiv.org/abs/1402.5903}{\tt arXiv:1402.5903}].

\bibitem{MGSDJ}
A.~Moradi~Marjaneh, V.~A.~Gani, D.~Saadatmand, S.~V.~Dmitriev, and K.~Javidan, \href{https://doi.org/10.1007/JHEP07(2017)028}{{JHEP} {\bf 07}, 028 (2017)} [\href{https://arxiv.org/abs/1704.08353}{\tt arXiv:1704.08353}].

\bibitem{GaLeLi}
V.~A.~Gani, V.~Lensky, and M.~A.~Lizunova, \href{https://doi.org/10.1007/JHEP08(2015)147}{{JHEP} {\bf 08}, 147 (2015)} [\href{https://arxiv.org/abs/1506.02313}{\tt arXiv:1506.02313}].

\bibitem{GaLeLiconf} V.~A.~Gani, V.~Lensky, M.~A.~Lizunova, and E.~V.~Mrozovskaya, \href{https://doi.org/10.1088/1742-6596/675/1/012019}{{J.~Phys.: Conf.~Ser.} {\bf 675}, 012019 (2016)}  [\href{https://arxiv.org/abs/1602.02636}{\tt arXiv:1602.02636}].

\bibitem{Radomskiy.JPCS.2017}
R.~V.~Radomskiy, E.~V.~Mrozovskaya, V.~A.~Gani, and I.~C.~Christov, \href{https://doi.org/10.1088/1742-6596/798/1/012087}{{\it J.~Phys.: Conf.~Ser.} {\bf 798}, 012087 (2017)} [\href{https://arxiv.org/abs/1611.05634}{\tt arXiv:1611.05634}].

\bibitem{Belendryasova.arXiv.2017}
E.~Belendryasova and V.~A.~Gani, , \href{https://arxiv.org/abs/1708.00403}{\tt arXiv:1708.00403}.

\bibitem{GaKuPRE}
V.~A.~Gani and A.~E.~Kudryavtsev, \href{https://doi.org/10.1103/PhysRevE.60.3305}{{Phys.~Rev.~E} {\bf 60}, 3305 (1999)} [\href{https://arxiv.org/abs/cond-mat/9809015}{\tt cond-mat/9809015}].

\bibitem{Gani.arXiv.2017.dsg}
V.~A.~Gani, A.~Moradi Marjaneh, A.~Askari, E.~Belendryasova, and D.~Saadatmand, \href{https://arxiv.org/abs/1711.01918}{\tt arXiv:1711.01918}.

\bibitem{Bazeia.arXiv.2017.sinh}
D.~Bazeia, E.~Belendryasova, and V.~A.~Gani, \href{https://arxiv.org/abs/1710.04993}{\tt arXiv:1710.04993}.

\bibitem{Bazeia.arXiv.2017.sinh.conf}
D.~Bazeia, E.~Belendryasova, and V.~A.~Gani, \href{https://arxiv.org/abs/1711.07788}{\tt arXiv:1711.07788}.

\bibitem{Bazeia.PRD.2002}
D.~Bazeia, L.~Losano, and J.~M.~C.~Malbouisson, \href{https://doi.org/10.1103/PhysRevD.66.101701} {{\it Phys.~Rev.} {\bf D 66}, 101701 (2002)} [\href{https://arxiv.org/abs/hep-th/0209027}{\tt hep-th/0209027}].

\bibitem{Bazeia.PRD.2006}
D.~Bazeia, M.~A.~Gonz\'alez Le\'on, L.~Losano, and J.~Mateos Guilarte, \href{https://doi.org/10.1103/PhysRevD.73.105008}{{Phys.~Rev.~D} {\bf 73}, 105008 (2006)} [\href{https://arxiv.org/abs/hep-th/0605127}{\tt hep-th/0605127}].

\bibitem{GaKsKu01}
V.~A.~Gani, V.~G.~Ksenzov, and A.~E.~Kudryavtsev, \href{https://doi.org/10.1134/S1063778810110104}{{Phys.~Atom.~Nucl.} {\bf 73}, 1889 (2010)} [{Yad.~Fiz.} {\bf 73}, 1940 (2010)] [\href{https://arxiv.org/abs/1001.3305}{\tt arXiv:1001.3305}].

\bibitem{GaKsKu02}
V.~A.~Gani, V.~G.~Ksenzov, and A.~E.~Kudryavtsev, \href{https://doi.org/10.1134/S1063778811050085}{{Phys.~Atom.~Nucl.} {\bf 74}, 771 (2011)} [{Yad.~Fiz.} {\bf 74}, 797 (2011)] [\href{https://arxiv.org/abs/1009.4370}{\tt arXiv:1009.4370}].

\bibitem{Bazeia.EPJC.2017}
D.~Bazeia and A.~Mohammadi, \href{https://doi.org/10.1140/epjc/s10052-017-4778-9}{{Eur.~Phys.~J.~C} {\bf 77}, 203 (2017)} [\href{https://arxiv.org/abs/1702.00891}{\tt arXiv:1702.00891}].

\bibitem{Lensky.JETP.2001}
V.~A.~Lensky, V.~A.~Gani, and A.~E.~Kudryavtsev,  \href{https://doi.org/10.1134/1.1420436}{{Sov.~Phys.~JETP} {\bf 93}, 677 (2001)} [{Zh.~Eksp.~Teor.~Fiz.} {\bf 120}, 778 (2001) ] [\href{https://arxiv.org/abs/hep-th/0104266}{\tt hep-th/0104266}].

\bibitem{Bazeia.MPLA.2002}
D.~Bazeia, W.~Freire, L.~Losano, and R.F.~Ribeiro, \href{https://doi.org/10.1142/S0217732302008435}{{\it Mod.~Phys.~Lett.} {\bf A 17}, 1945 (2002)} [\href{https://arxiv.org/abs/hep-th/0205305}{\tt hep-th/0205305}].

\bibitem{Kurochkin.CMMP.2004}
V.~A.~Gani, N.~B.~Konyukhova, S.~V.~Kurochkin, and V.~A.~Lensky, {Comput.~Math.~Math.~Phys.} {\bf 44}, 1968 (2004) [{Zh.~Vychisl.~Mat.~Mat.~Fiz.} {\bf 44}, 2069 (2004)] [\href{https://arxiv.org/abs/0710.2975}{\tt arXiv: 0710.2975}].

\bibitem{Bazeia.EPJC.2014}
D.~Bazeia, A.~S.~Lob\~ao Jr., L.~Losano, and R.~Menezes, \href{https://doi.org/10.1140/epjc/s10052-014-2755-0}{{ Eur.~Phys.~J.~C} {\bf 74}, 2755 (2014) } [\href{https://arxiv.org/abs/1312.1198}{\tt arXiv:1312.1198}].

\bibitem{GaLiRa}
V.~A.~Gani, M.~A.~Lizunova, and R.~V.~Radomskiy, \href{https://doi.org/10.1007/JHEP04(2016)043}{{JHEP} {\bf 04}, 043 (2016)} [\href{https://arxiv.org/abs/1601.07954}{\tt arXiv:1601.07954}].

\bibitem{GaLiRaconf}
V.~A.~Gani, M.~A.~Lizunova, and R.~V.~Radomskiy, \href{https://doi.org/10.1088/1742-6596/675/1/012020}{{J.~Phys.: Conf.~Ser.} {\bf 675}, 012020 (2016)} [\href{https://arxiv.org/abs/1602.04446}{\tt arXiv:1602.04446}].

\bibitem{Gani.YaF.1999}
V.~A.~Gani et al, {\it Phys.~Atom.~Nucl.} {\bf 62}, 895 (1999) [{\it Yad.~Fiz.} {\bf 62}, 956 (1999)] [\href{https://arxiv.org/abs/hep-ph/9712526}{\tt hep-ph/9712526}].

\bibitem{Gani.YaF.2001}
T.~I.~Belova, V.~A.~Gani, and A.~E.~Kudyavtsev, \href{https://doi.org/10.1134/1.1344952}{{\it Phys.~Atom.~Nucl.} {\bf 64}, 140 (2001)} [{\it Yad.~Fiz.} {\bf 64}, 143 (2001)] [\href{https://arxiv.org/abs/hep-ph/0003308}{\tt hep-ph/0003308}].

\bibitem{GaKu.SuSy.2001}
V.~A.~Gani and A.~E.~Kudryavtsev, \href{https://doi.org/10.1134/1.1423755}{{Phys.~Atom.~Nucl.} {\bf 64}, 2043 (2001)} [{Yad.~Fiz.} {\bf 64}, 2130 (2001)] [\href{https://arxiv.org/abs/hep-th/9904209}{\tt hep-th/9904209}, \href{https://arxiv.org/abs/hep-th/9912211}{\tt hep-th/9912211}].

\bibitem{GaKiRu}
V.~A.~Gani, A.~A.~Kirillov, and S.~G.~Rubin, \href{https://arxiv.org/abs/1704.03688}{\tt arXiv:1704.03688}.

\bibitem{GaKiRu.conf}
V.~A.~Gani, A.~A.~Kirillov, and S.~G.~Rubin, \href{https://arxiv.org/abs/1711.07700}{\tt arXiv:1711.07700}.

\end{thebibliography}
\end{document}